\documentstyle[12pt,epsfig]{article}
\def\subfigureA#1{
\leavevmode
\hbox{#1}
}
\def\MTfig#1#2#3#4{
   \begin{figure}[h]
   \begin{center}\mbox{\epsfig{figure=#2,height=#3cm}}\end{center}
   \caption{\footnotesize #4}
   \label{fig:#1}
   \end{figure}
      }
\newcommand{\m}{\medbreak}

\newcommand{\no}{\noindent}
\newcommand{\EQ}{\begin{equation}}
\newcommand{\eq}{\end{equation}}
\newcommand{\EQA}{\begin{eqnarray}}
\newcommand{\eqa}{\end{eqnarray}}

\newcommand{\AR}{\renewcommand {\arraystretch}{1.5}
\begin{array}{l}}
\newcommand{\bAR}{\renewcommand {\arraystretch}{2}
\begin{array}{l}}
\newcommand{\ARc}{\renewcommand {\arraystretch}{1.5}
\begin{array}{c}}
\newcommand{\bARc}{\renewcommand {\arraystretch}{2}
\begin{array}{c}}
\newcommand{\ar}{\end{array} \renewcommand {\arraystretch}{1}}

\newcommand{\eg}{\;=\;}

\def\plan{$\cal{P}\;$}
\def\planr{$\cal{P}_R\;$}

\def\planp{$\cal{P}$}
\def\planrp{$\cal{P}_R$}

\begin{document}
\begin{titlepage}
\vspace{0.2in}
\vspace*{0.5cm}
\begin{center}
{\large \bf Probing Dark Energy with Supernovae : Bias from the time evolution
of the equation of state}
\\

\m
\vspace*{0.8cm}
J.-M. Virey, P. Taxil\\
\vspace*{0.5cm}
Centre de Physique Th\'eorique,
Case 907,
F-13288 Marseille Cedex 9, France\\
and 
Universit\'e de Provence, Marseille, France\\
\vspace*{0.8cm}

A. Tilquin, A. Ealet, D. Fouchez, C. Tao\\
\vspace*{0.5cm}
Centre de Physique des Particules de Marseille,
Case 907\\
F-13288 Marseille Cedex 9, France
\vspace*{0.5cm}

{\bf Abstract} \\
\end{center}
Observation of thousands of type Ia supernovae should offer
the most direct approach to probe the dark energy content of the universe.
This will be undertaken by future large ground-based surveys followed by 
a space mission (SNAP/JDEM).
We address the
problem of extracting the cosmological parameters from the future data
in a model independent approach, with  minimal assumptions on
 the prior knowledge of some parameters.
We concentrate on the comparison
between a fiducial model and the fitting function and adress in particular
the effect of neglecting (or not)
the time evolution of the equation of state. 
We present a quantitative analysis of the bias which can be
introduced by the fitting procedure. Such bias cannot be ignored
as soon as the statistical errors from present data  are drastically improved.

\vfill
\begin{flushleft}
PACS Numbers : 98.80.Es, 98.80.Cq\\
Key-Words : cosmological parameters - supernovae
\m\no
Number of figures : 10\\

\m\no
March 2004\\
CPT-2004/P.009\\
CPPM-P-2004-01\\
\m\no
anonymous ftp or gopher : cpt.univ-mrs.fr

E-mails : virey@cpt.univ-mrs.fr, tilquin@cppm.in2p3.fr
\end{flushleft}
\end{titlepage}

   
\pagestyle{myheadings}

\section{Introduction}

\indent
\m

The general paradigm in cosmology is that we are living in a flat universe,
which is dominated by a nearly homogeneous component with negative pressure. 
This component
is often called Dark Energy (DE) and causes the expansion rate of the universe 
to accelerate.

The recent measurements of type Ia supernovae (thereafter denoted SN) are the most direct
evidence of the presence of this component \cite{SCP,Riess,Tonry,newSCP}.
It is also confirmed by the combination of results from the large-scale distribution of galaxies
\cite{Efst2DF} and  the most precise data on the cosmic microwave
background (CMB) from the Wilkinson Microwave Anisotropie
Probe (WMAP) \cite{Spergel}. 
A  recent combined analysis is presented in \cite{SDSSWMAP}.\\
Recently the detection of the late  Integrated Sachs-Wolfe (ISW) effect 
has reinforced the case for DE \cite{Scranton}.
All this is most frequently interpreted in the framework of the  
so-called "concordance Cosmology" \cite{Bahcall}.
\m

A fundamental problem is the identification of the
underlying nature of DE : cosmological constant, quintessence 
(for a review see \cite{Peebles}) or something else.
The most common way is to measure its equation of state (EoS) defined
as $w \eg p_X/\rho_X\,$ where $p_X\,$ is the pressure and $\rho_X\,$ the energy density of
the DE. The ratio of the DE density to the critical density will be
denoted $\Omega_X$ in a general model and $\Omega_{\Lambda}$ in the simplest
case of a Cosmological Constant ($w = -1$). $\Omega_M$ is the corresponding parameter
for (baryonic+cold dark) matter.

An ambitious SN program  is now on the way.  
Important pieces of information will be provided by large ground-based surveys 
such as the
Supernova Legacy Survey (SNLS) \cite{SNLS} and these investigations
will culminate with
a space mission as the Supernova Acceleration Probe (SNAP) instrument, 
part of the Joint Dark Energy Mission (JDEM), which aims at the 
discovery and follow-up of some 2000 SNIa per year in the redshift range $z = 0.2-1.7$ with
very precise magnitude measurements \cite{SNAPweb}.
Of course, the validity of the obtained precision
will not depend on the size of the SN sample only, but also on the ability
to control the systematic uncertainties at the same level
\cite{Kim}. These systematic errors are coming on one side
from the instrument itself (calibration, etc, ...) and also from the SN astrophysical
environment (evolution, lensing effects, ...). The SNAP collaboration proposes a global strategy
to control such effects at the percent level \cite{SNAPweb}. 
\m
In this context, it is necessary to analyze very carefully at which
precision level it will be possible to draw
any conclusions  from the expected rich amount of data.

Authors have used the present SN data together with simulated sets to  :
\begin{itemize}
\item evaluate the accuracy on the $\Omega_{\Lambda}$ parameter to validate the acceleration.
\item measure the DE through the equation of state with $w$ constant in redshift.
\item evidence a possible redshift dependence $w(z)$ with various parametrizations.
\item look at the model dependences and degeneracies and study the possible
strategies to break the latter.
\end{itemize}
 
There is a large consensus that the future SN data {\em alone} will
have difficulties to constrain an evolving equation of state and
to break model degeneracies. It is mandatory 
to have a prior knowledge of the values of some parameters.
In particular, a precise knowledge of $\Omega_M$ will be essential
if one hopes to pick out the $z$ dependence of $w$ (see e.g.
\cite{Maor1,Maor2,WA,HT,Goliath,Gerke,LH}) even in the simplest flat cosmology.
However, to use some pieces of information from other sources than SN,
it will be essential to combine the data in
a coherent way, that is under the same hypothesis, in particular on the
DE properties.
\m
On the other hand, some potential difficulties have been pointed out by
various authors. Indeed, most papers have been
mainly interested in predicting, for each cosmological parameter, the
errors around some fiducial value, close to the ones obtained from
the present data as 
$\Omega_M = 0.3, \Omega_X = 0.7, w = -1, dw/dz = 0$. 
In this procedure, the framework is a particular fiducial model and the chosen fitting function 
is usually  the one used to generate the data.
This strategy is valuable for a first estimate but is too restrictive to pin down the underlying
physics which is so far unknown. 

For example, Maor et al. \cite{Maor2} 
and Gerke and Efstathiou \cite{Gerke} have stressed the problems which arise
if the redshift dependence of the equation of state is neglected whereas this
dependence is present. The multiple integral relation 
between the luminosity distance and $w$ smears out information about
$w$ and its time-variation. 
Assuming in
the fitting function that $w$ is a constant, whereas it is not in the fiducial model,
can lead to a very wrong estimate of the "constant" or even "effective" $w$ value.
At the same time, the central value of  $\Omega_M$ (or $\Omega_X$) is
badly reconstructed. 
\\

It is unavoidable to get some ambiguities when trying to
fit a particular fiducial cosmology with the "wrong" model ! 
In the following, we will call this problem the "bias problem". 
For present SN data, as will be shown, this question is not a concern as the statistical errors 
on model parameters stay large due to the limited statistics. 
In the perspective of much richer SN samples, this bias problem cannot
be ignored.
 We address the problem in a more quantitative way than the authors of \cite{DiPietro} which have 
only considered some specific
classes of models. In this paper, we investigate the full parameter range, 
in a model independent way, to quantify where the 'bias' creates 
potential difficulties and where it can be ignored.
We also evaluate the impact of priors when only SN measurements are used.
We  get a handle on the possible 
bias introduced in the analysis by looking at the different hypothesis which are done at the fitting
function level. Results are presented in the parameter space
of the fiducial models we have simulated.

\m
This paper is organized as follows. In Section 2 we recall
the theoretical description of  the magnitude-redshift diagram for
supernovae and we present the
experimental situation. We identify the relevant parameters which enter as the parameters
of the fiducial models and we define the bias which could
be introduced by the fitting procedure.

In section 3, we focus on the
extraction of the EoS parameter $w_0$. 
We illustrate the confusion created by the bias on examples. Then we analyze the
impact of bias on simulated data : we consider future
statistical samples corresponding to SNAP, SNLS and also a sample corresponding to present data.
We try to find the safest way
for extracting an effective $w$ from the data with minimum
bias and minimum priors. 
Our summary and conclusions are given in Section 4.

\section{The experimental and theoretical framework}

\subsection{The experimental framework}

 The studies presented in this paper are performed with simulated supernovae 
samples with statistics equivalent to what we expect to have
in the future. We concentrate on three sets of data which simulate the statistical
power of the present and future data. We want to emphasize the importance of the future sample of
supernovae taken on ground which are limited by the
systematics inherent to this approach but are statistically one order 
of magnitude greater than the present sample.  
\begin{itemize}
\item We reproduce the published data of Perlmutter et al. from \cite{newSCP}. We take directly the effective
magnitude of the sample 3 of this paper, which corresponds to 60 SNIa, corrected from K corrections, extinction and stretch
(the values are given in Table 3 of \cite{newSCP}). The errors on magnitude vary in the range [0.16-0.22].
\item We simulate some future data coming from ground survey as the large SNLS survey at CFHT \cite{cfht}.
 This survey has started in 2003 and the estimation after 5 years of running is to register a sample
of 700 identified SNIa in the redshift range $0.3<z<1$. 
We simulate a sample as reported in Table~\ref{tab:stat} in agreement with the expected rate of \cite{cfht}. 
We assume a magnitude dispersion of 0.15 for each supernova 
and constant in redshift after all corrections.  
\item We simulate data from a future space mission like SNAP, which plans to
discover around 2000 identified SNIa, at redshift  0.2$<z<$1.7 with very precise photometry and spectroscopy. 
The SN distribution is given in 
Table~\ref{tab:stat}(from\cite{Kim}). The magnitude dispersion is assumed to be constant and independent
of the redshift at 0.15 for all SN's after correction. On some studies, we include
 the effect of adding a constant and uncorrelated in redshift systematic
error of 0.02 on the magnitude.
\end{itemize}
A set of 300 very well calibrated SnIa at redshift $< 0.1$ should be measured by the incoming 
SN factory project \cite{SNfactory}.  This sample is needed
to normalize the Hubble diagram and will be called in the following the "Nearby" sample. \\
A "SNAP" ("SNLS") sample means in this paper a simulation of the statistics expected from
the SNAP like mission ("SNLS" survey) combined with the 300 nearby SN's.
To make our point clear,  we do not consider in the presented studies the experimental systematic errors, 
unless otherwise specified.

{\small 
\begin{table}[htb]
\begin{center}
\begin{tabular}{|l|c|c|c|c|c|c|c|c|c|c|c|c|c|c|c|c|}
\hline
   z & 0.2 & 0.3 & 0.4 & 0.5 & 0.6 & 0.7 & 0.8 & 0.9 & 1.0 & 1.1 & 1.2 & 1.3 & 1.4 & 1.5 & 1.6 & 1.7 \\ \hline
SNLS & -   & 44  & 56  & 80  &  96 & 100 & 104 & 108 & -   & -   & -   & -   & -   & -   & - & - \\ 
SNAP & 35  & 64  &  95 & 124 & 150 & 171 & 183 & 179 & 170 & 155 & 142 & 130 & 119 &  107& 94 & 80 \\
  \hline
\end{tabular}
\caption{\sl number of simulated supernovae by bin of 0.1 in redshift for SNLS and SNAP respectively.}  
\label{tab:stat}
\end{center}
\end{table}
}

\subsection{The theoretical framework}

In the standard Friedmann-Roberston-Walker metric, the apparent magnitude of 
astrophysical objects can be expressed as a function of the
luminosity distance :

\EQ\label{mag}
m(z) \eg 5\,log_{10}(D_L) + M_B  - 5 \,log_{10}(H_0/c) + 25 \eg M_s + 5 \,log_{10}(D_L)
\eq
\noindent
where $M_B$ is the absolute magnitude of SNIa,
$M_S$ may be considered as a normalization parameter
 and 
$D_L(z)\equiv (H_0/c)\;d_L(z)$ is the \emph{$H_0$-independent} luminosity distance to an object 
at redshift $z$.  
It is related to the comoving distance $r(z)$ by \\
$D_L(z) \eg (1+z)r(z)$,
where 
\begin{eqnarray}
    r(z)&=&\left\{
    \begin{array}{ll}
      \frac{1}{\sqrt{-\Omega_k}}\sin(\sqrt{-\Omega_k}\,J) , &
      \Omega_k<0\\
      \,J , & \Omega_k=0\\
      \frac{1}{\sqrt{\Omega_k}}\sinh(\sqrt{\Omega_k}\,J) , &
      \Omega_k>0\\
    \end{array}
    \right. \\
   {\rm with \;} \;\Omega_k&=&1-\Omega_m-\Omega_X \equiv 1-\Omega_T,\\
    J&=& \int_0^z\,\frac{H_0 }{H(z')}dz' 
\end{eqnarray}

\no 
and 
\EQ
  \left(\frac{H(z)}{H_0}\right) ^2 \eg (1+z)^3\,\Omega_{m}+
    {\rho_X(z)\over \rho_X(0)} \, \Omega_{X}+(1+z)^2\,\Omega_{k}, \\
\eq
with 
\EQ
\label{rho}
{\rho_X(z)\over \rho_X(0)} \eg \exp \left[ 3\int_0^z\,\left(1+w(z')\right)\, d\,\ln (1+z') \right] 
\eq

\m 
Note that we have neglected
the radiation component $\Omega_R $.

\subsubsection{Choice of a fitting function}

A fitting function is an expression for $H(z)$ in terms of a  number
of parameters. $H(z)$ is usually expressed 
in term of the cosmological parameters contained in 
Eq.(5)\cite{Maor1,Maor2,WA,HT,Goliath,Gerke,LH} but other 
possibilities relying on geometrical parametrizations
or arbitrary ansatz
 have been proposed \cite{geo}. In this paper, we adopt the first approach
where the cosmological parameters have a direct physical interpretation
within General Relativity. With such analysis, the constraints are aimed to be model
independent and can be used for investigation of a large variety of DE models. 
However, we have to choose a parametrization for the time (or redshift)
variation of the EoS.

The current parametrizations of $w(z)$ which can be found in the literature
are the linear one \cite{WA}: $w(z) = w_0 + w_1z$; the one advocated by Linder \cite{LinderPRL}
with a better
asymptotic behavior : $w(z) = w_0 + w_az/(1+z)$; and also 
$w(z) = w_0 - \alpha \, \ln (1+z)$ from \cite{Gerke} 
which is advocated to describe a large sample of 
quintessence models characterized by a weakly varying EoS. 
We  choose the linear one which allows
to compare most easily to previous works. 
In this case, Eq.(\ref{rho}) becomes:
\EQ
\rho_X(z)\eg \rho_X(0) \, e^{3w_1z} \, (1+z)^{3(1+w_0-w_1)}
\eq
Nevertheless, our results are essentially independent of the choice of this parametrization 
 since
the redshift range of the SN data is limited to $z<2$ and 
the sensitivity of the data on $w(z)$
concerns mainly the values taken by $w$ at relatively low redshifts $z\simeq 0.2-0.4$
(the "sweet spot" of \cite{HT,WA,LH}) where all parametrizations reduce roughly to the linear one.

Finally, note that models with non-trivial variations of the EoS
$w(z)$, like the pseudo-goldstone model of ref.\cite{Frieman}, cannot be
described by such parametrizations. Consequently, our analysis does not apply to this kind of models.\\

Therefore, five parameters have to be fitted in the most general procedure: \\
$M_s,\Omega_M, \Omega_X, w_0, w_1$.

\subsubsection{Choice of a fiducial model}

We define
 fiducial models which depend on
the  parameters $M_s^F,\Omega_M^F, \Omega_X^F,w_0^F, w_1^F$.\\

In order to keep this paper clear, we concentrate on the case where the paradigm 
$\Omega_T=1$ is verified by the fiducial models. This value
 is in agreement with the inflationary paradigm and it is
supported by the analysis of the rich amount of data from the CMB \cite{Spergel}.
$\Omega_X^F$ is then no longer free ($\Omega_X^F=1-\Omega_M^F$) and, if not stated otherwise, 
the parameters $M_s^F$ and $\Omega_M^F$ are fixed at: $M_s^F=-3.6$ and $\Omega_M^F=0.3$.

Our conclusions will not depend on variations of the $M_s^F$ parameter.
The variation due to different $\Omega_M^F $ values has been investigated.
If $\Omega_T \neq 1$, it appears that our 
results can change significantly. Analysis of non-flat models will
be presented elsewhere \cite{enpreparation}.\\
 

To focus on the dark energy measurement, we propose to scan a large 
variety of fiducial models as a function of the couple
 ($w_0^F,w_1^F$). We have scanned the plane ($w_0^F,w_1^F$) 
(hereafter denoted by \planp ) for the values $-2<w_0^F<0$ and $-2<w_1^F<2$. 
We pay particular attention to the reduced plane \planr associated to the
ranges $-1<w_0^F<0$ and $-1<w_1^F<1$, which represents a reasonable class
of theoretical models. 
Indeed, in most of the models available in the literature,
like the quintessence models \cite{Peebles}, the weak energy condition
 $w_0^F>-1$ is satisfied (see e.g. \cite{Frampton}). They also possess in general a relatively
weak $z$ dependence for the EoS \cite{Gerke,WA}.
However, we consider the full plane \plan in order to include in our analysis more
exotic models which violate the weak energy condition or which are described by
a modified Friedmann equation or by a modified theory of gravity (we refer
to \cite{Nesseris} for a recent list of models and references).\\

\subsubsection{The fitting method}

To analyse the (simulated) data,  a minimization procedure has been used \cite{kosmoshow}.
A standard Fisher matrix approach allows a fast estimate of  the parameter errors. 
This method is however limited as it does not 
yield the central values of the fitted parameters.
Then we adopt, unless specified, a minimization procedure based on  a least square method.
The least square estimators are determined by the minimum of the 
$ \chi^2 = ({\bf m} - M(z,\Omega, w)^T {\bf V}^{-1} ({\bf m} - M(z,\Omega, w))$, where
${\bf m} = (m_1...m_n)$ is the vector
of magnitude measurements, $M(z,\Omega, w)$ the corresponding vector of predicted values
 and {\bf V} the covariance matrix of the measured magnitudes.
 The error on cosmological parameters is estimated
at the minimum by using the first order error propagation technique: $ {\bf U = A.V.A^T} $ where {\bf U}
is the error matrix on the cosmological parameters and {\bf A} the Jacobian of the transformation. 

A full 5-parameter fit (5-fit) of the real or simulated data gives
 the central values and errors for the five parameters.
In the following, the normalization parameter $M_s$ is left free in the fit, unless specified,
as its correlation with the cosmological parameters is strong \cite{Goliath}. 
To constrain accurately 
 this parameter, which is fixed in most of the papers, we 
use the very low redshift simulated ``Nearby'' sample of 300 SN. \\





\subsection{Choice of a fitting procedure}

There are different ways to choose a fitting procedure. If one only wants to 
minimize the errors on the fitting parameters, a simple Fisher approach should be sufficient
and should give conclusions depending on the initial fit hypothesis. 
Nevertheless, as we have already emphasized, this approach is fiducial model 
dependent and can lead to some bias that we 
want to quantify. Then, after a presentation of different Fisher results, we will expose 
our strategy to understand and explore the bias introduced by some fitting approach.

\subsubsection{The Fisher approach}
To perform a Fisher matrix analysis, a fiducial model is chosen and  fixed. We take the "concordance model" version of the simplest flat $\Lambda$CDM model :\\
$$
\Omega_M^F=0.3,\; \Omega_X^F=0.7,\; w_0^F=-1,\; w_1^F=0 \;{\rm with} \;M_s^F=-3.6$$ 
We call this model '$\Lambda$' in the following. 

Each {\em fitting procedure} is defined by a particular choice of the parameters to be fitted:
reducing the number of parameters will improve the parameter errors
in the fitting function. Priors are introduced by fixing a parameter at a predefined value or inside a
given range. 
\begin{table}[htbp]
\begin{center}
\caption{\protect\footnotesize Statistical errors obtained with a Fisher matrix analysis on the cosmological
parameters for various 
fitting procedures with the SNLS and SNAP data repectively. The fiducial model is a cosmological constant with
$ M_s^F=-3.6,\;
\Omega_M^F=0.3,\; \Omega_X^F=0.7,\; w_0^F=-1,\; w_1^F=0$. The weak (strong) $\Omega_M$ prior corresponds
to the constraint $\Omega_M=0.3\pm 0.1$ ($\Omega_M=0.3\pm 0.01$). The labels 5-fit, 4-fit and 3-fit
corresponds to the fitting procedures 5-fit($M_s,\Omega_M, \Omega_X, w_0, w_1$), 4-fit($M_s,\Omega_M, w_0, w_1$)
and 3-fit($M_s,\Omega_M, w_0$), respectively. $\sigma(\Omega_X)$ has been omitted from the table since
$\Omega_X=1-\Omega_M$ is no longer a free parameter except for the 5-fit where
$\sigma(\Omega_X)$ is very large. $\sigma(M_S)$ has been omitted from the table since
its value, roughly 1\% , changes weakly for the various fitting procedures thanks to the inclusion
of the Nearby SN sample.}
\m
\vspace*{0.5cm}
\label{tab:fisher}
\begin{tabular}{ c|c|c||c|c|c||c|c|c}
\hline 
 Fit &  $\Omega_M$ prior & assumptions &  $\sigma(\Omega_M)$
& $\sigma(w_0)$   &  $\sigma(w_1)$ &  $\sigma(\Omega_M)$
& $\sigma(w_0)$   &  $\sigma(w_1)$  \\ 
  & & & SNAP& SNAP& SNAP & SNLS& SNLS& SNLS \\
\hline \hline
5-fit & no & no & 1.51 & 4.15 & 11.07 & 19.10 & 63.37 & $>$100 \\
4-fit & no  &$\Omega_T = 1$ & 0.14 & 0.11 & 1.31 & 1.08 & 1.21 & 7.94  \\
4-fit & weak &$\Omega_T = 1$ & 0.082 & 0.076 & 0.77 & 0.10 & 0.161 & 0.89   \\
4-fit & strong &$\Omega_T = 1$ & 0.01 & 0.055 & 0.174 & 0.01 & 0.117 & 0.511 \\
3-fit & no &$\Omega_T = 1$,$w_1=0$ & 0.016 & 0.063 & / & 0.069 & 0.19 & / \\
3-fit & weak &$\Omega_T = 1$,$w_1=0$ & 0.016 & 0.063 & / & 0.057 & 0.158 & /   \\
3-fit & strong &$\Omega_T = 1$,$w_1=0$ & 0.008 & 0.040 & / & 0.01 & 0.047 & / \\
\hline
\end{tabular}
\end{center}
\end{table}

Table~\ref{tab:fisher} gives the errors obtained with the Fisher matrix analysis
for various fitting procedures  with the SNLS and SNAP simulated data. 
The numbers are in agreement
with the analysis of various authors \cite{Maor1,Maor2,WA,HT,Goliath,Gerke,LH}. \\
 Figure~\ref{fig:fisher} gives the contours obtained in the plane ($\Omega_M, w_0$) at 68.3 \% CL 
with the SNAP statistics, for this fiducial model
within the same fitting procedures. \\
As can be seen from the first line of Table~\ref{tab:fisher}  
and from the black ellipse in Fig.~\ref{fig:fisher},
a complete 5 parameter-fit (5-fit)($M_s,\Omega_M, \Omega_X, w_0, w_1$), with no assumptions 
on the values of the cosmological parameters - 
including no constraint on the flatness of the universe- yields too
large errors on  the cosmological parameters to set any definite conclusions even with
the high statistics sample expected for the SNAP+Nearby data.\\ 
 Fig.~\ref{fig:fisher}  shows clearly that the reduction
of the number of fitted parameters or/and the strengthening of the $\Omega_M$ prior,
reduces the contours.\\

\MTfig{fisher}{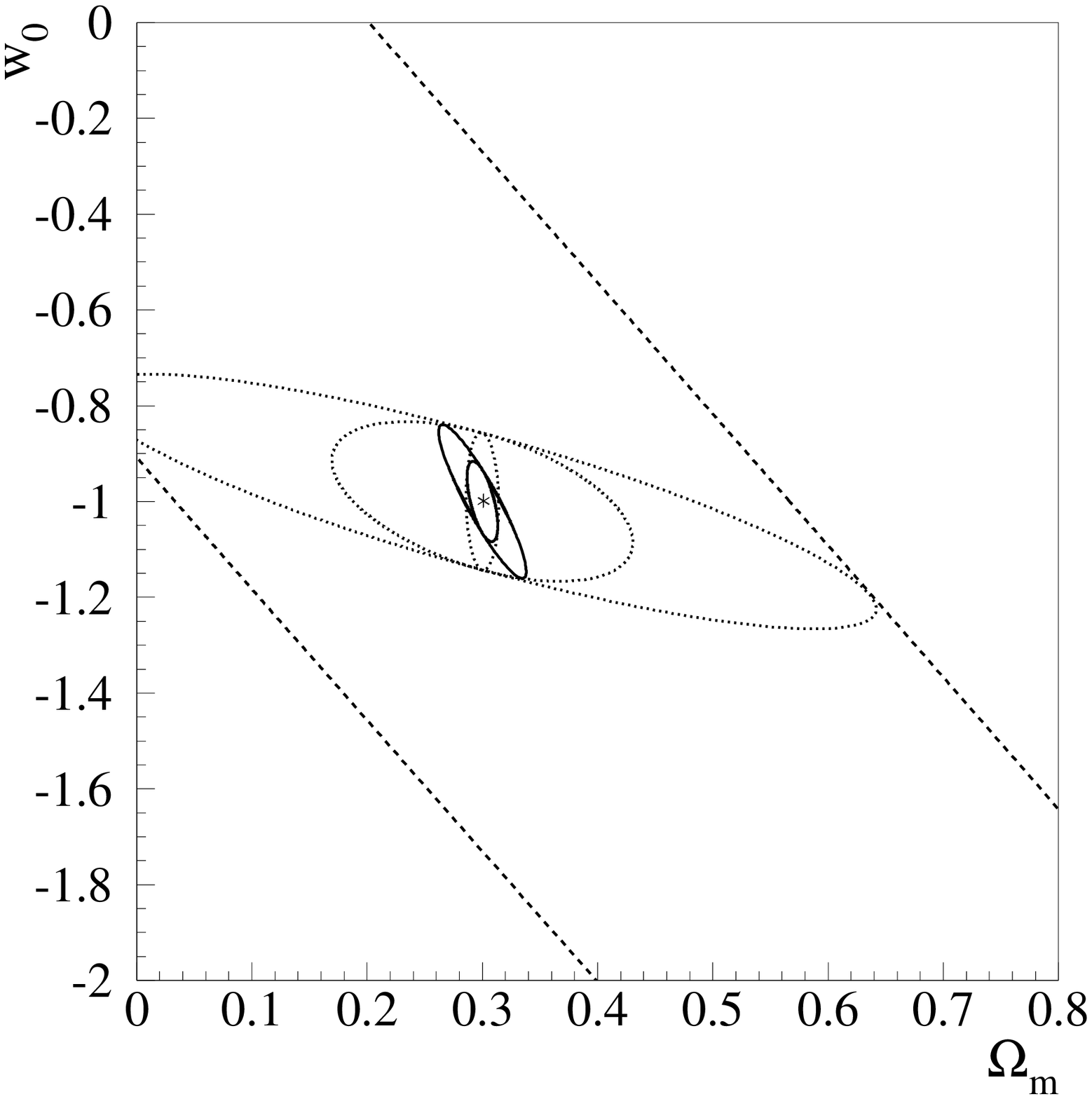}{7.}{
Fisher contours in the plane ($\Omega_M, w_0$) at 68,3 \% CL for a cosmological constant
 as the fiducial model.
The large dashed contour corresponds to the
5-fit($M_s,\Omega_M, \Omega_X, w_0, w_1$). The three dotted contours correspond to the 
4-fit($M_s,\Omega_M, w_0, w_1$) with no prior (largest), with a weak $\Omega_M$ prior (i.e. $\Omega_M=0.3\pm 0.1$)
and with a strong $\Omega_M$ prior (i.e. $\Omega_M=0.3\pm 0.01$, smallest  ellipse).
The solid contours correspond to the 
3-fit($M_s,\Omega_M, w_0$) with a weak $\Omega_M$ prior (largest)
and with a strong $\Omega_M$ prior (smallest). For this fitting procedure, there is almost no difference
between the cases "no" and "weak" prior for $\Omega_M$.}

 The second line of Table~\ref{tab:fisher} presents errors with the paradigm
of a perfectly flat universe ($\Omega_T = 1$)
 and the 5-fit reduces to a 4-fit ($M_s,\Omega_M, w_0, w_1$).
It appears that for the SNAP statistics, one gets 
better estimates of $w_0$ and $\Omega_M$ but the error on $w_1$ is always large.
For SNLS, even the error on $w_0$ is large. 
\m
Adding some prior knowledge of $\Omega_M$ improves greatly the situation: 
the same 4-fit ($M_s,\Omega_M, w_0, w_1$) for SNAP and SNLS with a
weak prior on $\Omega_M$ ($0.2 < \Omega_M < 0.4$) yields
 a good estimation of $w_0$  but still a very large error for $w_1$ (see line 3 of Table~\ref{tab:fisher} ).
Only a strong prior on $\Omega_M$ ($\Omega_M = 0.3 \pm 0.01$)
 really improves the situation for $w_1$ as can be seen from line 4 of Table~\ref{tab:fisher}.
These conclusions are in agreement with other published results 
\cite{Maor1,Maor2,WA,HT,Goliath,Gerke,LH}.\\

One can conclude that the determination of
a possible redshift dependence of the EoS remains the most difficult task.
All authors agree on the following statement:
to get a good precision on the parameter governing the redshift evolution of the 
EoS, a fit with a strong
prior on $\Omega_M$ ("strong" means at the percent level) is needed within the $\Omega_T = 1$ paradigm.

\subsubsection{The bias problem}
\m
Several comments on the previous approach are in order :
\begin{itemize}
 \item[-] The setting of strong priors inside the fit should be taken with caution. \\
First of all, the cosmological parameter $\Omega_M$ 
is far from being measured at the percent level \cite{Blanchard}. 
So, this kind of prior cannot be applied blindly, even if there is hope that this parameter might
be well measured by the  time  SNAP provides data.

\item[-] Even in a future context, some doubts can be raised about 
the relevance of very precise priors \cite{Maor1,Maor2} and their use has to be set with some cautions : 
 all the $\Omega$'s should be obtained
under the same hypothesis (astrophysical, experimental, DE properties) and correlations 
among fitted parameters have to be taken properly into account for the combined
statistical analysis to be relevant.
\end{itemize}

In this paper, we attempt to extract the informations from SN data, therefore avoiding
the (consistency) problems encountered when constraining $\Omega_M$ with external
data.

The results from Table~\ref{tab:fisher} and Fig.~\ref{fig:fisher} 
lead us to leave aside the determination
of the redshift dependence of the EoS and to consider $w_0$ as an "effective" constant
$w^{eff}$ EoS parameter. 
Then we concentrate on the best strategy
to extract $w_0$ from present or future SN data with minimal assumptions on priors :
we use the paradigm $\Omega_T=1$
either with no priors or with weak priors on $\Omega_M$.
 We  focus on a 3-fit ($M_s, \Omega_M, w_0$) or on a 4-fit ($M_s,\Omega_M, w_0, w_1$)
where we know already that the precision on $w_1$ will be low.
\\

A 3-fit should always provide better constraints than a 4-fit.
Table~\ref{tab:fisher} shows, however, that the reduction of the errors
is more important for $\Omega_M$ than for $w_0$ and since the $w_0$ error is 
not strongly improved, the 
use of the 4-fit seems the best strategy for extracting $w_0$ \cite{LH,Kim}.
We want to point out that this conclusion is only valid for the $\Lambda$ fiducial model.
In some other fiducial models,  the 3-fit can increase
considerably the constraints on $w_0$ with respect to the 4-fit: Figure~\ref{fig:bias} 
 gives the expected contours in the plane ($\Omega_M$,$w_0$)  for the two fiducial models
($w_0^F=-2$,$w_1^F=0$) and ($w_0^F=-1$,$w_1^F=0.2$). 
The best constraints on $\Omega_M$ and $w_0$ are given by the 3-fit.\\

\MTfig{bias}{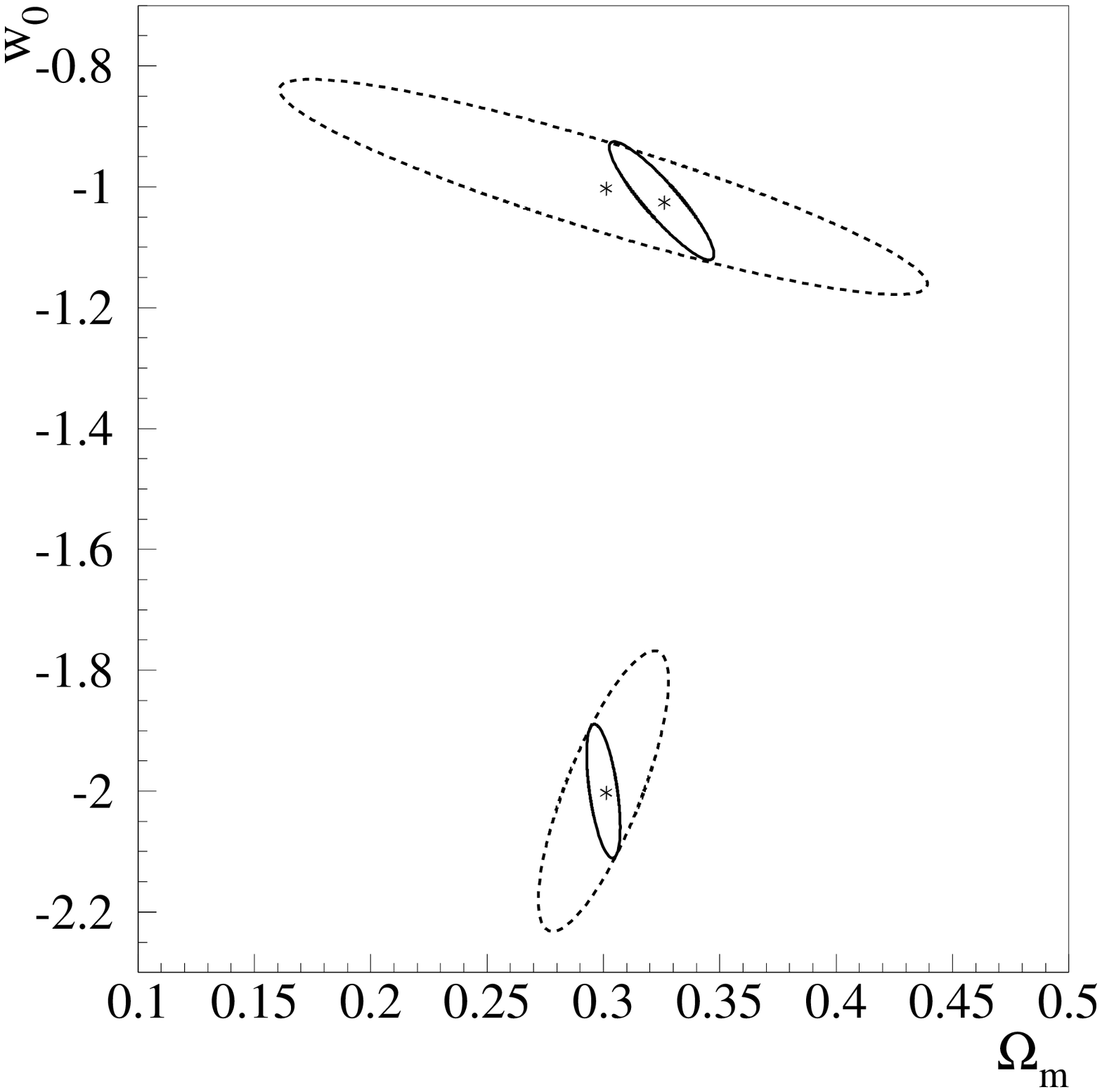}{7.}{
Contours in the plane ($\Omega_M, w_0$) at 68,3 \% CL for the fiducial models 
($w_0^F=-2$, $w_1^F=0$)
and ($w_0^F=-1$, $w_1^F=0.2$). 
The large dashed (small solid) contours correspond to the 
4-fit (3-fit). A weak $\Omega_M$ prior has been used.}

On the other hand, doing a complete minimization procedure shows that the central values
given by the 3-fit can be far away from the fiducial values. 
This fact has been pointed out by several authors  \cite{Maor1,Maor2,Gerke,DiPietro}.
This happens when the fitting hypothesis are not verified by the fiducial model, 
namely a constant EoS in the 3-fit and a varying one in the fiducial model.
In this case {\em bias
are introduced on the estimate of the cosmological parameters}.
This bias problem is due to the degeneracy of the fitting function among the various
cosmological parameters \cite{Maor1,Maor2}. A wrong hypothesis on one parameter is compensated
in a non trivial way by the value of the others.
We know also from the various Fisher analysis presented 
in the litterature \cite{HT,Goliath,LH,Kim} that the 
{\em errors} depend on the central
values of the cosmological parameters. Since these last ones are biased,
their errors are also biased \cite{LH,Kim}.\\ 
Then the question of choosing between the 3-fit and the 4-fit is not so obvious and appears to be fiducial model dependent. \\

In the next section we present in detail a study of this bias by exploring 
the range of fiducial models which is affected, in order to conclude on the validity of a 3-fit.\\
Let us point out that other bias would be introduced by 
a departure from the $\Omega_T=1$ paradigm in the fiducial model or if the central value
of the $\Omega_M$ prior used in the fitting procedure is not the same as the fiducial one.
These bias will be studied in a future paper \cite{enpreparation}.

\section{How to extract $w_0$ : a quantitative analysis of the validity of a 3-fit}

To be able to choose the best strategy for extracting $w_0$, we want to test the relevance of the 
fitting procedure. We have performed for that purpose a complete study of the bias introduced by the different 
fitting procedures. After a description
of the problem through a short illustration,  
we present the results of a full scan of the ($w_0^F,w_1^F$) plane for SNAP, SNLS
and a sample corresponding to present data.
 The scanning is done by fixing the fitting procedure and varying the
various parameters of the fiducial model ; we perform a 3-fit and 
compare it to the result of the 4-fit where $w_1$ is left free.

\m
\subsection{Illustration of the bias effect}

Let us start with some illustrations of the bias introduced by the 3-fit procedure.
We choose three different models and look at the central values of the parameters obtained from the fit:\\

\no $\bullet$ The one of Maor et al. \cite{Maor2} ($w_0^F=-0.7$, $w_1^F=0.8$).
The 3-fit applied to this model clearly provides some erroneous results :
 $\Omega_M = 0.62 \pm 0.013 $ and $w_0 = -1.548 \pm 0.194$.
Note that our central values are slightly different from the ones
of \cite{Maor2}, this is due to small differences between the chosen SN samples.

 Figure~\ref{fig:maor}  shows the two elliptical contours from the 3-fit and the 4-fit :
the 4-fit ellipse is centered on the fiducial values
as expected whereas the 3-fit ellipse is very far away. This is due to the 
large chosen $w_1^F$ value and this behavior has been already pointed out
by Maor et al. (see Fig. 5 of ref. \cite{Maor2}).\\
\MTfig{maor}{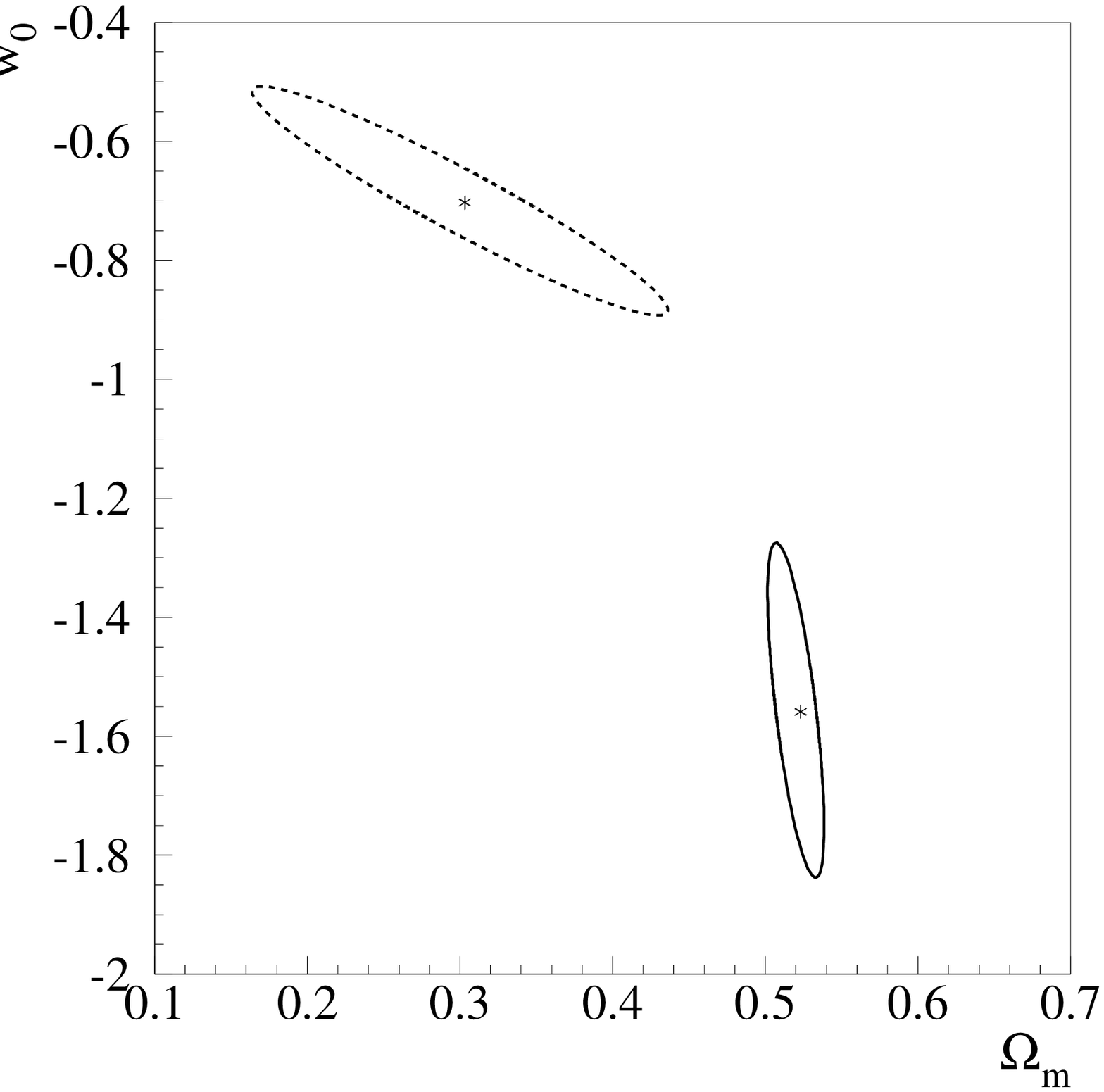}{7.}{
Contours in the plane ($\Omega_M, w_0$) at 68,3 \% CL for the fiducial models 
($w_0^F=-0.7$, $w_1^F=0.8$)(Maor et al. model\cite{Maor2}).
The dashed (solid) contour corresponds to the 
4-fit (3-fit) contour. A weak $\Omega_M$ prior has been used.
We can observe the large differences in the central values coming from
the two 3-fit and 4-fit fitting procedures.}

\no $\bullet$ If we take the model
($w_0^F=-1$, $w_1^F=0.2$), which has a ``small'' $w_1^F$ value, it leads to a smaller bias :
 we get
 $\Omega_M = 0.324 \pm 0.016 $ and $w_0 = -1.021 \pm 0.067$. 
We see that there is a bias of the order of 1.5$\sigma$ for
$\Omega_M$, and a small bias within the statistical error for $w_0$.\\
The central values (crosses) displayed in  Figure~\ref{fig:bias} of the ellipses for the 3-fit and the 4-fit are 
different but the 3-fit ellipse is included in the 4-fit one, indicating a small bias on one parameter
($w_0$) and a not too big bias on the second ($\Omega_M$).\\

\no $\bullet$ Finally, 
with the model ($w_0^F=-2$, $w_1^F=0$) which has no $z$ dependence
of the EoS, there is no bias
and both ellipses are centered on the same point (see Figure~\ref{fig:bias}).\\

Note that there is no significant bias for the normalization
parameter $M_S$ in the three cases.

\m

\subsection{Validity zones for the 3-fit}

To quantify the bias, we scan the full plane \plan ($w_0^F,w_1^F$) using the procedure 3-fit,
 assuming a flat universe. 
We first define different zones in this plane
where the effects of the bias can be quantitatively estimated. \\

The plane \plan is divided in three distinct zones :
\begin{itemize}

\item The Non Converging Zone (NCZ) where the fit has some detectable problems: the fit is
bad either because $\chi^2>3n$ ($n$ being the number of degrees of freedom), 
or because the error on one of the fitted parameter is above 1. 
We also reject 
fits yielding some $\Omega_M$ values far away from current expectations, namely we require
$0.1-\sigma(\Omega_M)<\Omega_M<1.+\sigma(\Omega_M)$.\\
This zone is shaded in our plots. None of the
fiducial models located in this NCZ can be fitted with the procedure 3-fit.
Therefore, with real data, the fit procedure will be excluded.

\item The Biased Zone (BZ) is the part of \plan where  the fit converges perfectly, 
 but where we know
from the simulation that we are far from the fiducial values:
We quantify this bias as the difference between the central value of a fitted parameter $\cal{O}$
and the fiducial value ${\cal{O}^F}$. Namely the "bias" is defined as  
\EQ
B({\cal{O}}) \eg |{\cal{O}}^{F}-\cal{O}|
\eq 
In general, $B \neq 0$ and we define the biased zone as the part of \plan  where 
$B({\cal{O}}) > \sigma (\cal{O})$
 where $ \sigma (\cal{O})$ is the statistical error on $\cal{O}$. 
\m
The important point is  that this zone is undetectable with real data.
It is the region we want to identify and to minimize 
through an appropriate choice of the fitting procedure.  

\item The Validity Zone (VZ) is the remaining part of \plan  
where the fit converges perfectly and where $B({\cal{O}}) < \sigma (\cal{O})$.  
If we require this condition for the full set of fitted parameters together,
 we can define a ``full'' validity zone
(full VZ).

Consequently, the contours separating the BZ and the VZ in the plane \plan 
correspond to the intrinsic limitation of the fitting procedure which is considered.\\
\end{itemize}

We now apply this strategy of analysis to the SNAP and SNLS simulated data. The impact of the bias
on the present SN statistical sample is also presented.\\

\subsection{Fitting procedures for SNAP}
 \subsubsection{Determination of the 3-fit validity zones }

\MTfig{snap}{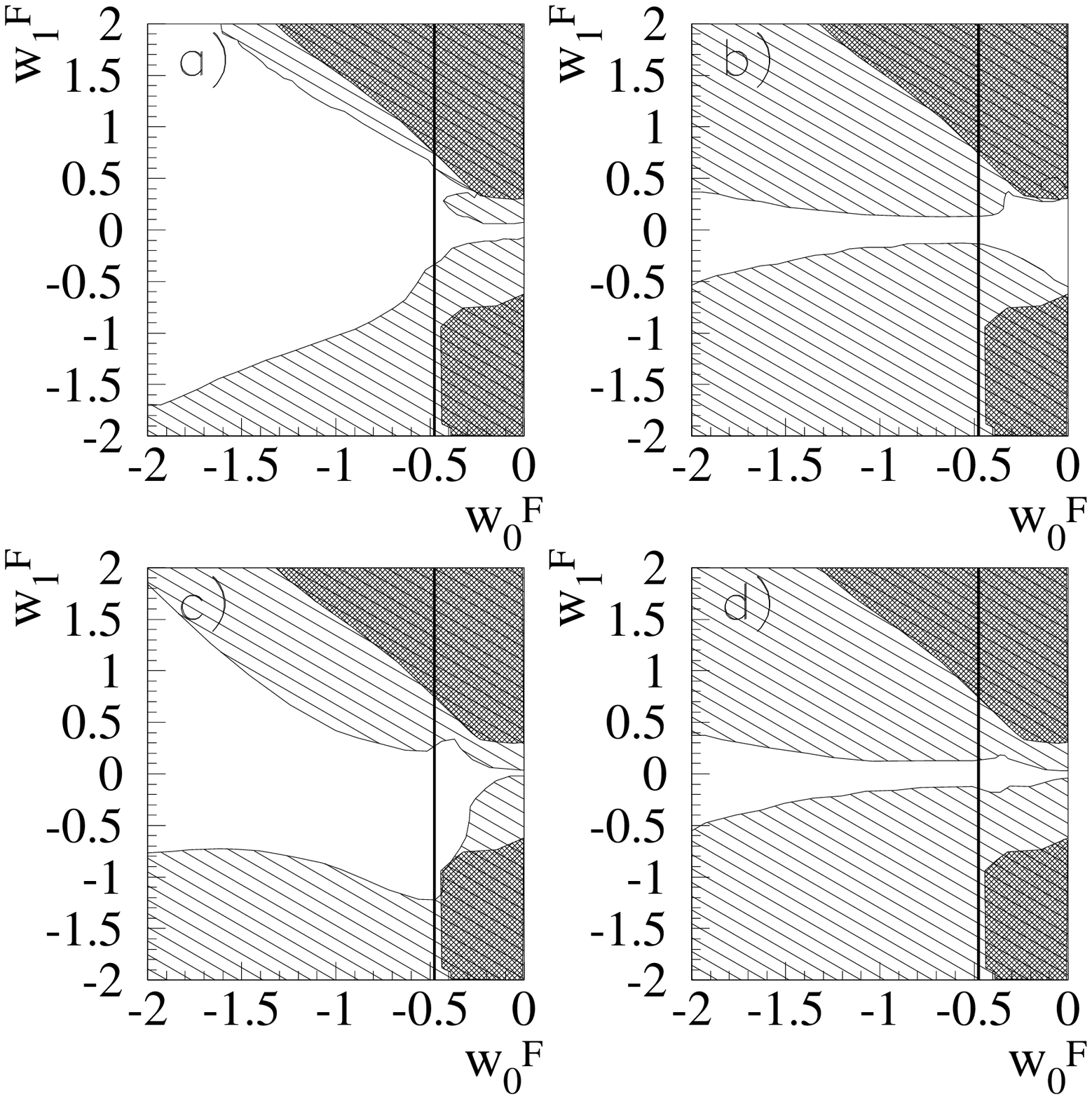}{10.}{Validity Zones(white), Biased Zones (hatched) and Non Converging Zones (black)
for a SNAP like data set 
for $M_S$ only (a), $\Omega_M$ (b),
$w_0$ (c) and all the parameters together (d). 
The procedure is a 3-fit with a weak $\Omega_M$ prior.
The vertical line at $w^F_0=-0.48$ separates decelerating models
from accelerating ones.}

We apply  this method to determine the validity zones using the 3-fit on the SNAP
simulated data. A weak prior on $\Omega_M$, $\Omega_M=0.3\pm 0.1$, has been used.\\
Figure~\ref{fig:snap}a,b,c,d give the different validity zones for
$M_s$ (Figure~\ref{fig:snap}a), $\Omega_M$ (Figure~\ref{fig:snap}b), $w_0$ (Figure~\ref{fig:snap}c)
and for the three parameters taken together (Figure~\ref{fig:snap}d).\\

In all the figures, the line $w^F_1=0$ corresponds to the actual unbiased fiducial models.\\
One can see that the Biased Zone is always limited by the NCZ. 
The line $w^F_0= -1/(3 \Omega_X^F) \simeq -0.48$ separates
the models which correspond to an accelerating universe today 
($w^F_0<-0.48$)  from the decelerating ones
 ($w^F_0>-0.48$).  \\

Looking at
the validity zone for each  parameter individually :
\begin{itemize}
\itemsep 4pt

\item  $M_s$ (Figure~\ref{fig:snap}a) is
 essentially unbiased for models which correspond to acceleration 
($w^F_0 < -0.48$) except for large negative values of $w_1^F$. We notice that $M_s$ is always measured with an error better than the percent thanks to the ``Nearby'' sample.
Therefore it is easy to get a bias greater than this error. However this bias has no real
consequence on the determination of the other parameters.

\item $\Omega_M$ (Figure~\ref{fig:snap}b), conversely, is unbiased only for
a small band around the line $w^F_1=0$. It means that for most of DE models
with varying EoS, the fitted $\Omega_M$ is strongly biased in this fitting procedure.

\item $w_0$ (Figure~\ref{fig:snap}c)  is valid in a 
large part of the 
reduced plane \planrp, which corresponds to the present acceleration region ($w^F_0<-0.48$).
Otherwise we fall in the Biased Zone which, for instance, contains
 the previously quoted model of Maor et al.\cite{Maor2}.
Outside \planrp , for large $\mid w_1^F \mid$ values, we are
in the Biased Zone.
Consequently, one has to take some cautions when using the  
3-fit for extracting $w_0$.
\\
\end{itemize}

The full Validity Zone is shown in Fig.~\ref{fig:snap}d ans is driven mainly by
the bias on the $\Omega_M$ parameter. 
One can conclude that
the 3-fit is  valid on all parameters together for fiducial models which 
have a small dependence in redshift: when $w^F_0> -1$, the validity is for $|w^F_1|<0.2$, otherwise it increases to
 $|w^F_1|<0.4$.  \\

One has to emphasize that the experimental error will also contain 
systematic errors. These latter have been estimated for SNAP \cite{Kim}.
We have simulated this case and the result
is given on Fig.~\ref{fig:statplussyst}, where we have included a constant 
uncorrelated systematic error
on each magnitude measurement. 
The consequence of increasing the error is an increase of the Validity Zones.  
Compared to Fig.~\ref{fig:snap}c, 
the VZ for $w_0$ is substantially enlarged. Note that the NCZ is reduced in the same time.
The VZ for $M_s$ follows the same behaviour, the one of $\Omega_M$ being
much less affected.

\MTfig{statplussyst}{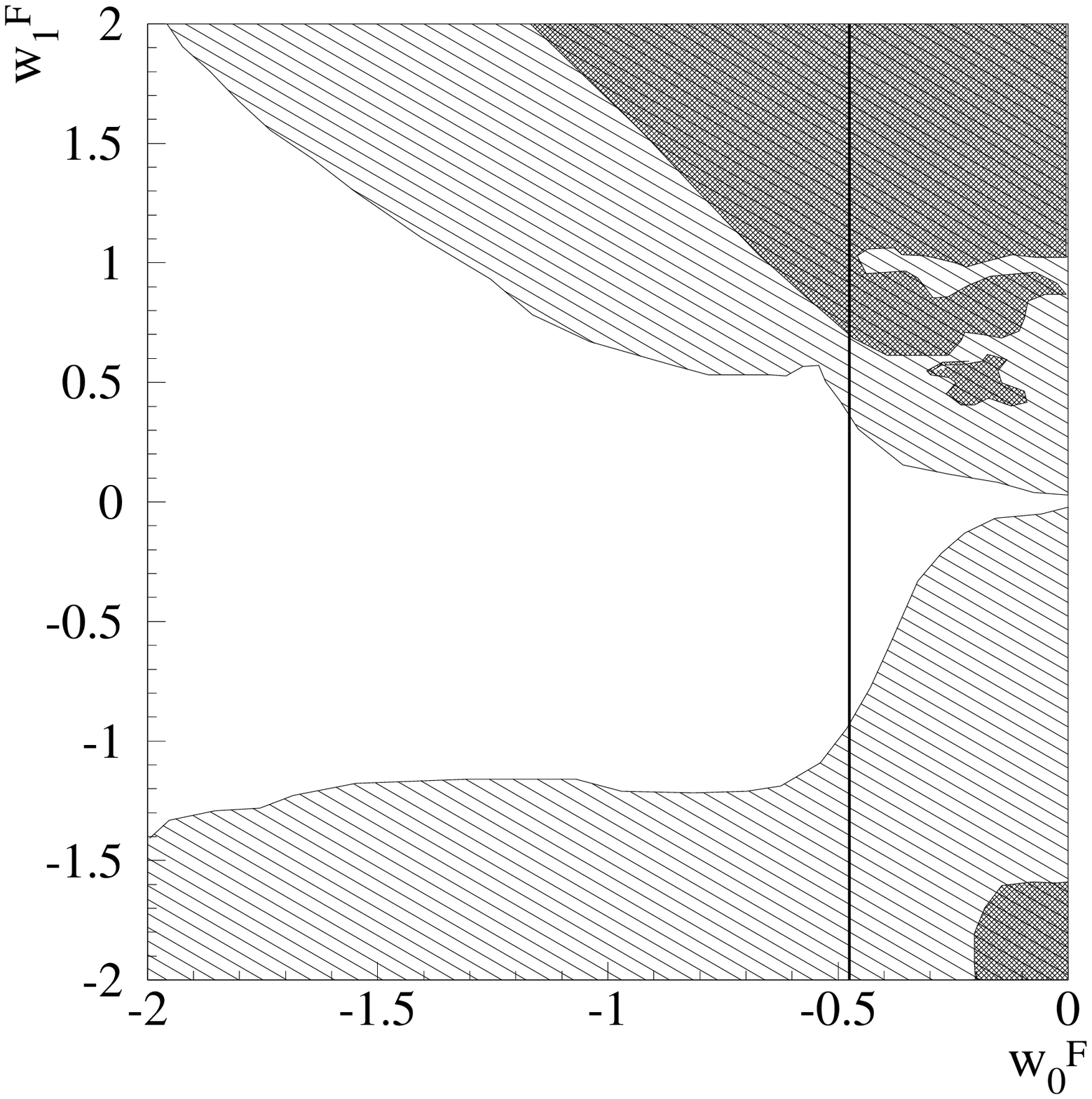}{8.}{Same as Fig.4c
for
$w_0$ in the case where we add an uncorrelated and constant in redshift
systematic error of 0.02 on the magnitudes.}

We have also looked at the impact of the $\Omega_M$ prior on our results: 

$\bullet$  When there is no prior, the results
are very comparable (but slightly worse) to the weak prior case described above.

\begin{figure}[ht]
\begin{tabular}[t]{c c}
\centerline{\subfigureA{\psfig{file={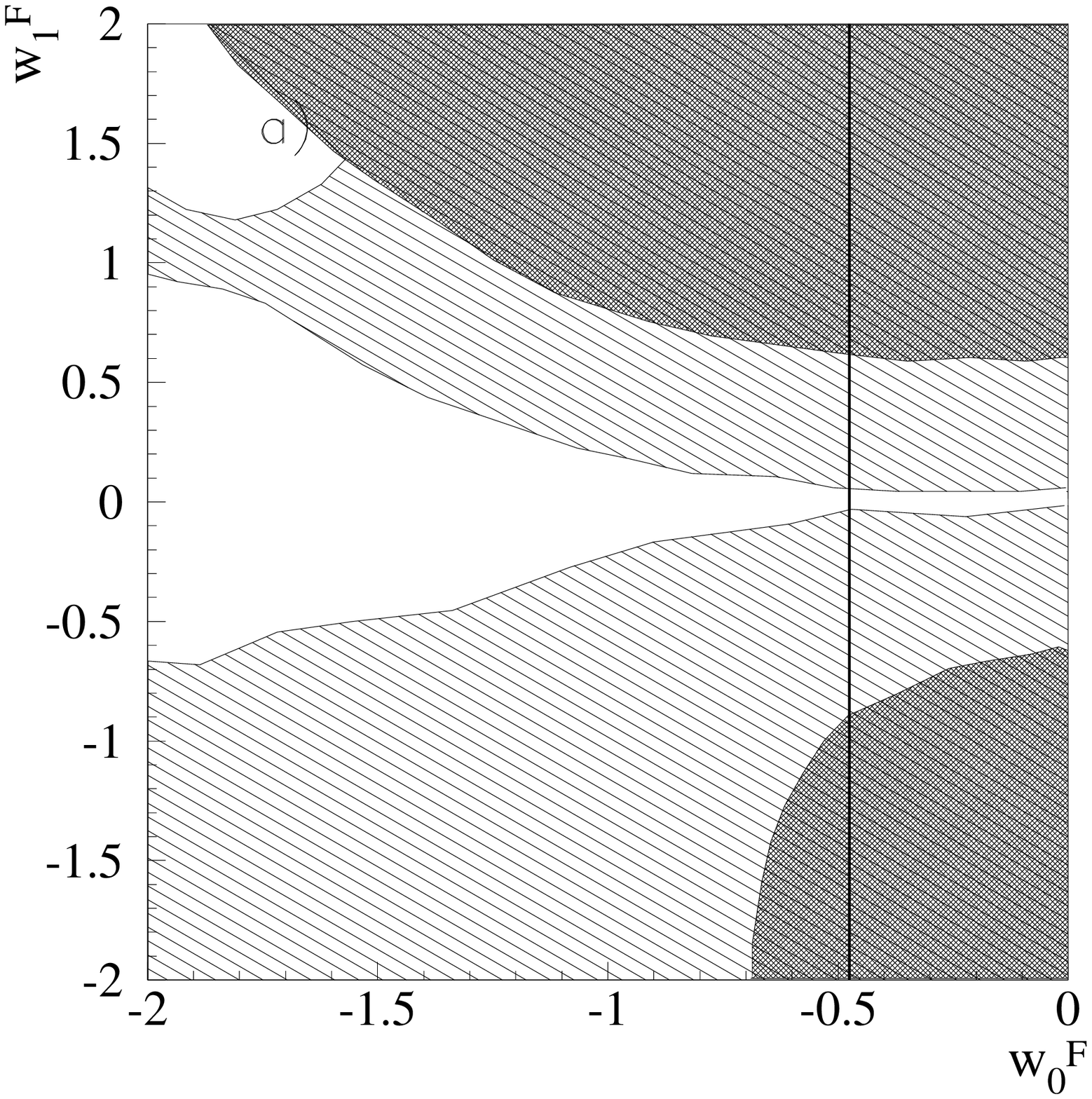},height=8truecm}}
\subfigureA{\psfig{file={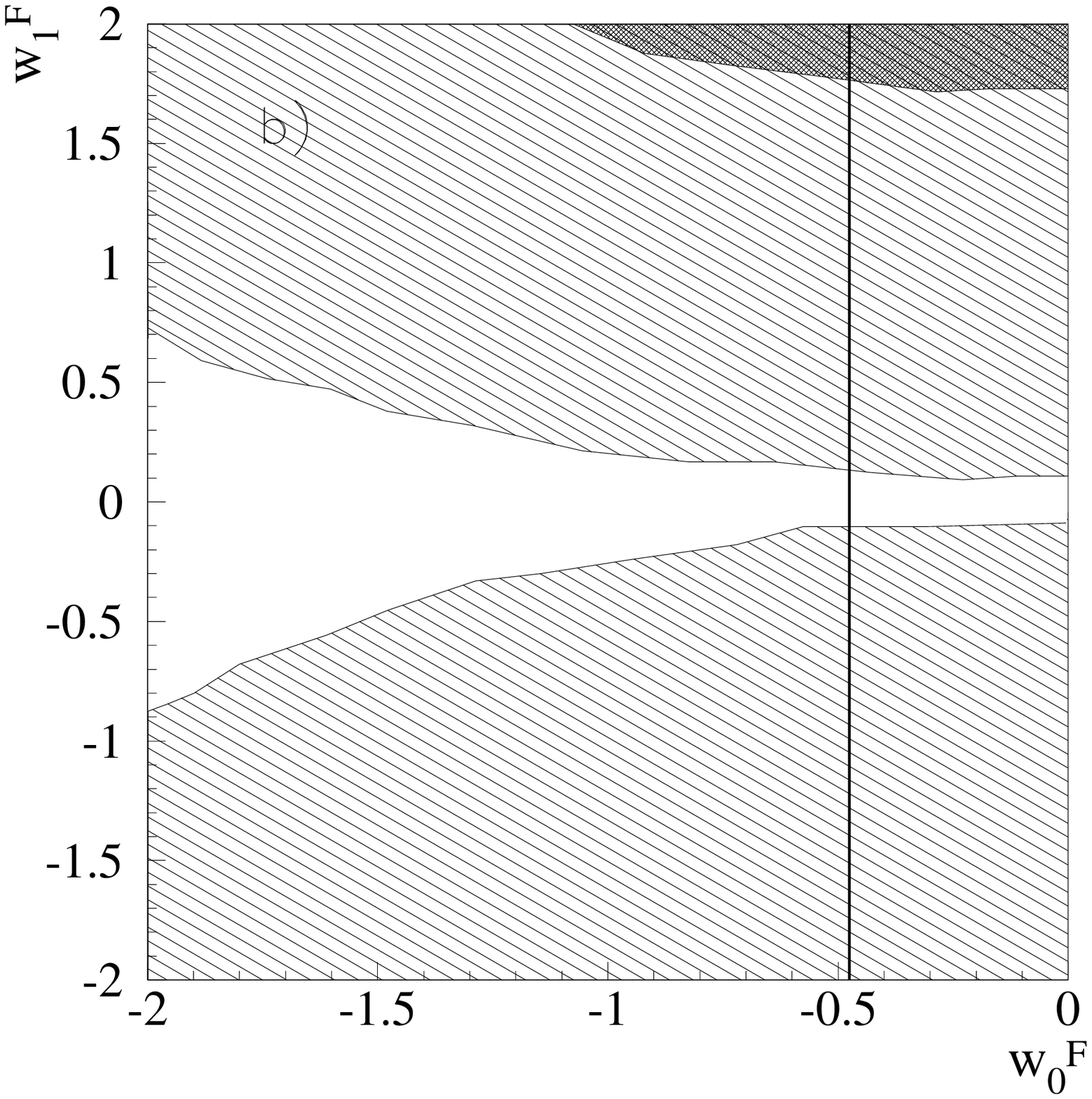},height=8truecm}}}
\end{tabular} 
\vspace{-1cm}
\caption{\footnotesize Same as Fig.4c
for $w_0$ in the case of a strong $\Omega_M$ prior
a) : SNAP, b) : SNLS.}
    \label{fig:prior} 
\end{figure}

$\bullet$ When a strong prior is used (e.g., $\sigma (\Omega_M )=0.01$), the error on the parameters
are reduced but the VZ also. Fig.~\ref{fig:prior}a gives the VZ for $w_0$ with this prior.
Due to the strong prior on $\Omega_M$, all the bias effect is reported on the
$w_0$ parameter which is strongly biased as shown on Figure~\ref{fig:prior}a. 
Looking more closely at the reduced \planrp, one sees that the VZ is limited to the line
$w^F_1=0$. Therefore we lose the fact that $w_0$ is in general well fitted by this fitting procedure.\\

The conclusion is that the fitting procedure 3-fit used in the case
of a high SN statistics experiment like SNAP, can be useful to constrain $w_0$
for a large part of the plane \planp, 
which corresponds to accelerating models, only when no prior or a weak prior on $\Omega_M$
is used. The other cosmological parameters ($\Omega_M$ especially) 
are strongly biased, and strengthening
the $\Omega_M$ prior worsens the situation.
Concerning DE models which lead to a present deceleration, it appears clearly
from the preceding figures, that the cosmological parameters are biased even if the
time dependence of the EoS is weak. 


\subsubsection{Comparison between the 3-fit and the 4-fit}

\m
\MTfig{errorsnap}{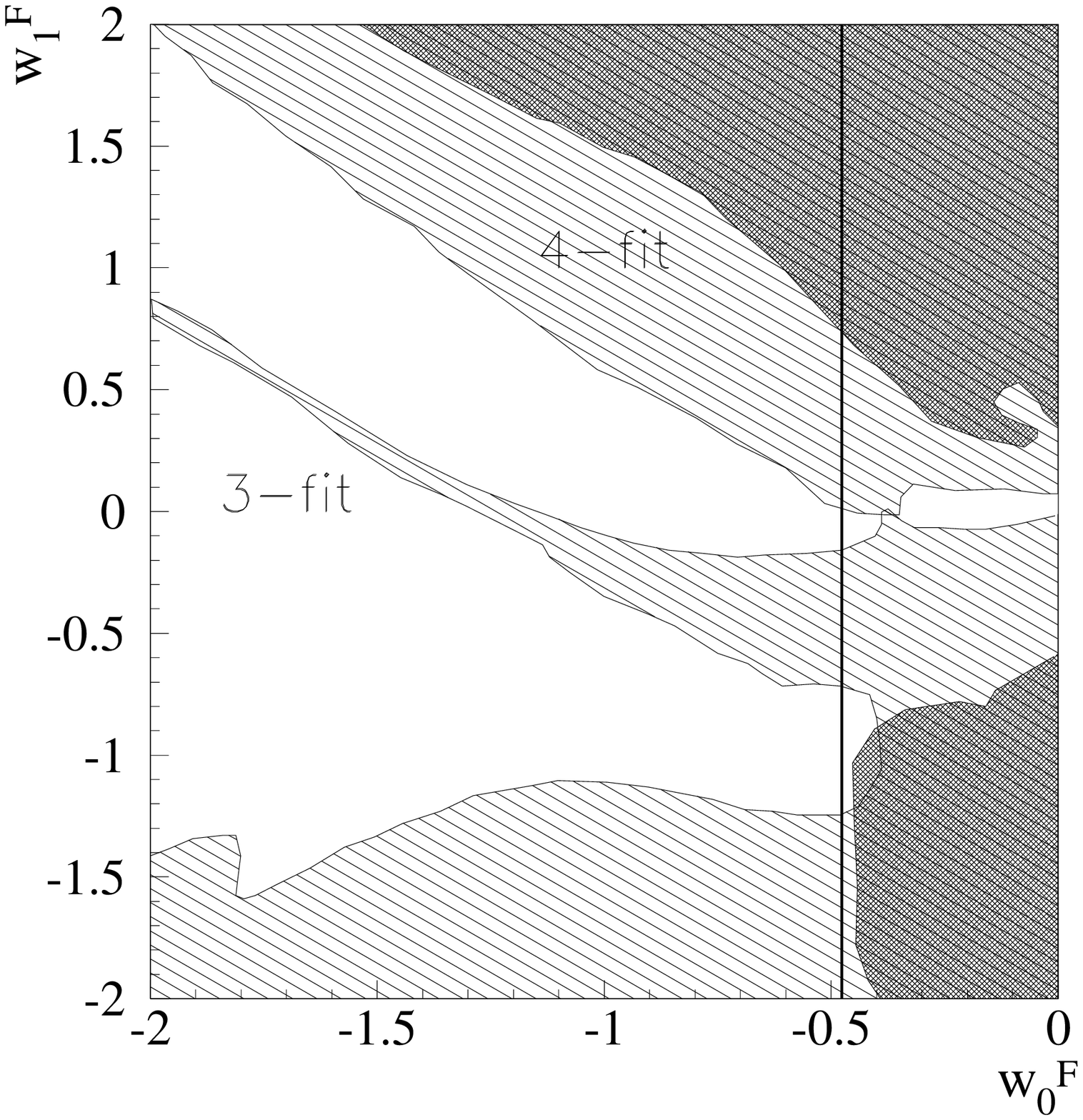}{8.}{
$E(w_0, {\rm 3-fit})- \sigma (w_0, {\rm 4-fit})$ for the SNAP data
 and a weak $\Omega_M$ prior. The white part corresponds to a positive difference, the hatched part
to a negative difference, the black region corresponds to the Non Converging zones.}

We have shown previously that the 3-fit procedure  
may be useful to extract $w_0$ even in the presence of the bias due to the hypothesis $w_1=0$.
The simplest 4-fit procedure (with no constraint on $w_1$)
allows the extraction of $w_0$ 
without any
bias since $w_1$ is included in the fit, but with an increase of the statistical errors on each parameter. 
We adress now a comparison
of the errors obtained on $w_0$ with these two different fitting procedures. \\

To combine the bias error and the statistical error, we choose a fit quality estimator of
the procedure as  $E(w_0)=\sqrt{\sigma^2(w_0) + B^2(w_0)}$ which reduces
to $\sigma(w_0)$ for the 4-fit.\\
Figure~\ref{fig:errorsnap} compares the two procedures by displaying the  difference of errors 
$E(w_0, {\rm 3-fit})- \sigma (w_0, {\rm 4-fit})$ with a weak prior on $\Omega_M$.
The 3-fit is preferred, with this estimator, in a large part of the plane \plan 
which corresponds to models with present acceleration.

The two black areas correspond to the NCZ of both fitting procedures.
Therefore, one cannot conclude for a preference for the 4-fit
in this zone. 
The two hatched zones, corresponding to large values of $|w_1^F|$, indicate
a preference for the 4-fit. This is due to the bias on $w_0$ inherent to the 3-fit.
%
There is also a narrow hatched band where the 4-fit
is still preferred although the bias is small. This appears to be a zone where the correlation between $w_0$ and $w_1$ is small
and then, the errors from the 4-fit and the 3-fit are similar, the 4-fit being
slightly better (less than 2\%).

In the white regions, the two parameters
$w_0$ and $w_1$ are strongly correlated and then the 3-fit has clearly a smaller error.

So, we can conclude that with a weak, thus conservative, prior on $\Omega_M$
the 3-fit is in general better than the 4-fit to extract $w_0$ for accelerating models.

\m

\subsection{Analysis for SNLS}
\subsubsection{Determination of the 3-fit validity zones }

A similar analysis has been performed for a simulated sample
corresponding to the SNLS-like statistics.\\

 The various Validity Zones obtained for SNLS with
a weak $\Omega_M$ prior are shown in Figures~\ref{fig:snls}a,b,c,d  \\

\MTfig{snls}{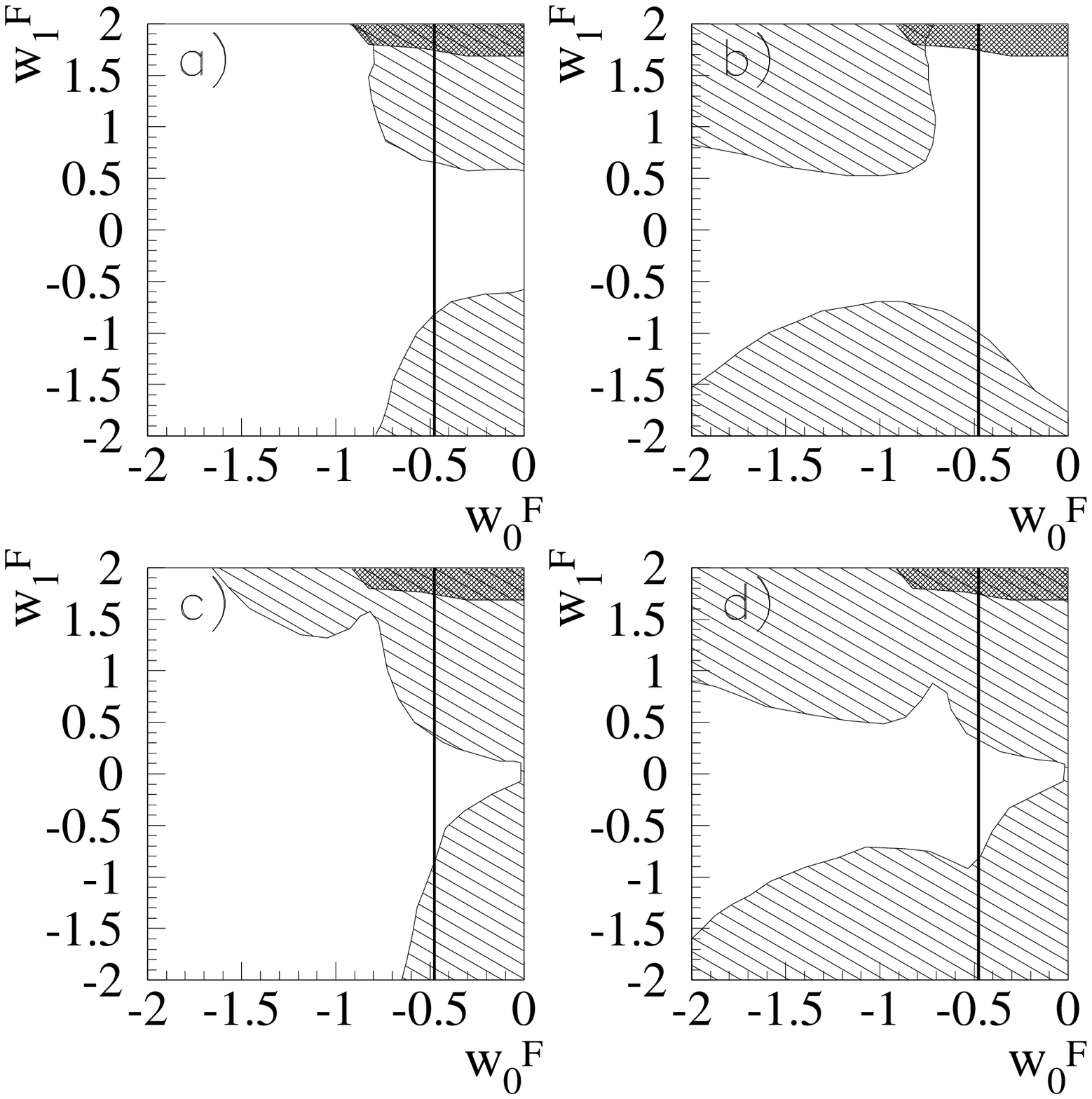}{10.}{Validity Zones (white), Biased Zones (hatched) and Non Converging Zones (black) 
for a SNLS like data set 
for $M_S$ only (a), $\Omega_M$ (b) and
$w_0$ (c) and all the parameters together (d). 
The procedure is a 3-fit with a weak $\Omega_M$ prior.}

We deduce from this figure that :
\begin{itemize}

\item 
 $M_S$ is almost unbiased  for accelerating models 
and also for decelerating ones provided the redshift 
dependence is small ($|w_1^F|<0.6$) (Fig.~\ref{fig:snls}a).
\item  $\Omega_M$ is not biased in most of
the reduded plane \planr and it is biased outside this plane if
$|w_1^F|>0.6$ (Fig.~\ref{fig:snls}b).
\item 
 $w_0$ is not biased for an important part of the plane corresponding to acceleration ($w_0<-0.48$) (Fig.~\ref{fig:snls}c).
\item The 3-fit is a good fitting procedure for reconstructing {\em all} cosmological
parameters simultaneously for accelerating models ($w_0<-0.48$) 
when the redshift dependence is small $|w_1^F|<0.6$ (see Fig.~\ref{fig:snls}d).
Otherwise the result is biased.
\item The use of a strong prior on $\Omega_M$ worsens the situation : $w_0$ is biased
for most of the plane \plan as shown on Fig.~\ref{fig:prior}b. 

\end{itemize}

Comparing these results with the SNAP case shows that the conclusions are quite similar but 
the validity zones for SNLS
are much larger than for SNAP. Due to the lower statistics our criterion $B({\cal{O}})<\sigma({\cal{O}})$ is less constraining. Note that the NCZ zones are strongly
reduced as expected with the lower statistics.\\

\subsubsection{Comparison between the 3-fit and the 4-fit}

We perform the same  comparison of the errors on $w_0$ obtained from the 
two procedures 3-fit and 4-fit,
still with a weak prior on $\Omega_M$.\\ 

\MTfig{errorsnls}{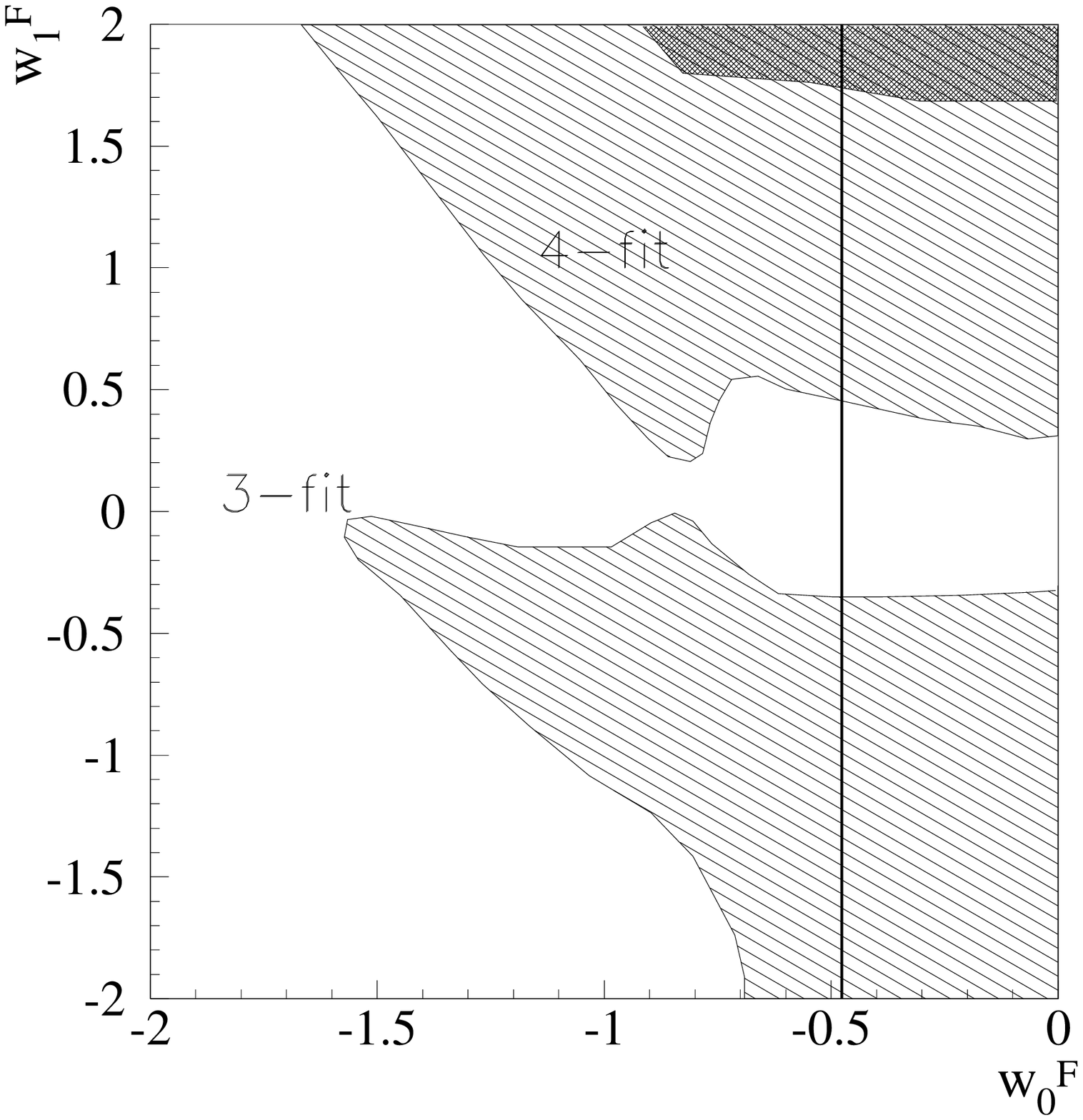}{8.}{$E(w_0, {\rm 3-fit})- \sigma (w_0, {\rm 4-fit})$  on $w_0$  for a SNLS like
 data set and a weak $\Omega_M$ prior. The white part corresponds to a positive difference, the dashed part
to a negative difference, the black region corresponds to the Non Converging Zones.}

Figure~\ref{fig:errorsnls} shows the  difference of errors between the 3-fit
and the 4-fit.
The 3-fit is preferred when the bias is small or when the two parameters $w_0$ and $w_1$ are strongly correlated.
 When $w_1^F<0$ the hatched region in the accelerating zone has similar
origin and properties as the narrow hatched zone of Fig. 7, namely that both fitting 
procedures provide in fact very comparable errors.
\\

\subsection{Analysis for a statistical sample corresponding to the SCP data}

We can define the validity zones for the statistics of the present SN sample
from the SCP collaboration \cite{newSCP} introduced in Section 2.1. 

\begin{figure}[ht]
\begin{tabular}[t]{c c}
\centerline{\subfigureA{\psfig{file={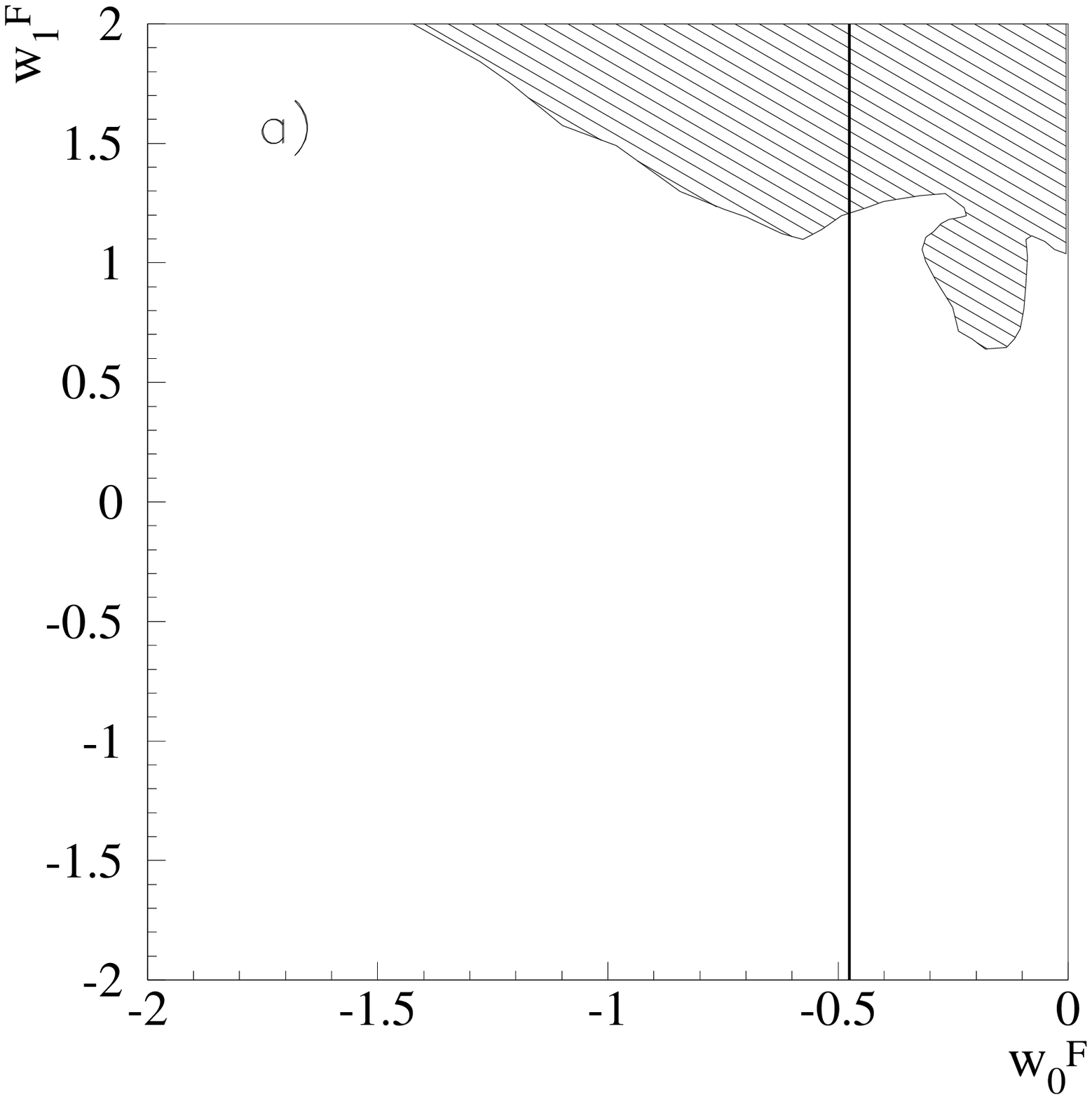},height=8truecm}}
\subfigureA{\psfig{file={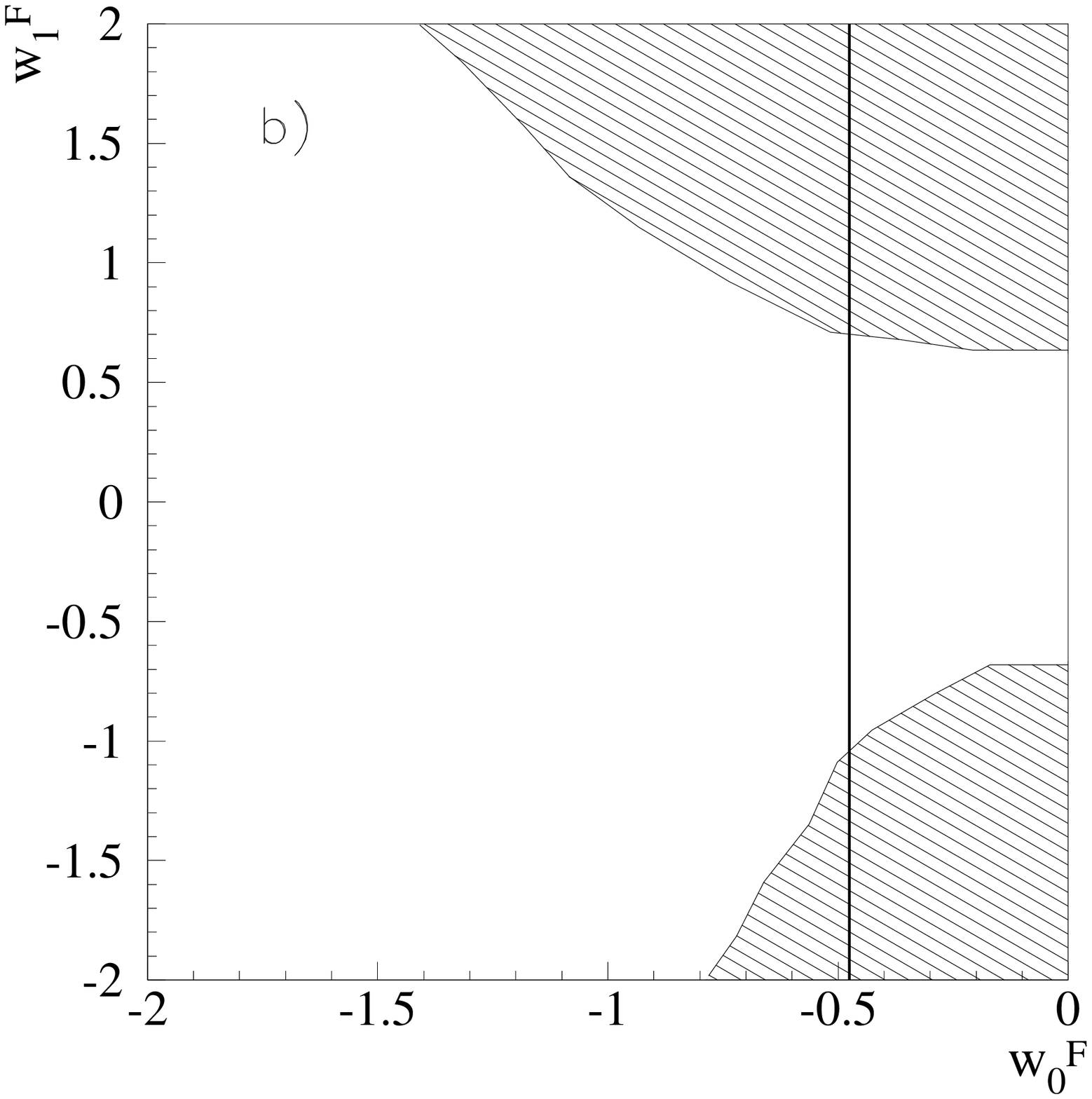},height=8truecm}}}
\end{tabular} 
\vspace{-1cm}
\caption{\footnotesize Full Validity Zones and Biased Zones 
for a SCP like data set 
 with a) no prior b) a weak $\Omega_M$ prior. The NCZ are not plotted due to
the large errors on $\Omega_M$ and $w_0$.}
    \label{fig:scp} 
\end{figure}

Figure~\ref{fig:scp}a  gives
the full Validity Zone obtained without any prior on $\Omega_M$.
The large size of this zone shows that the results are not biased (at $1\sigma$)
for most of the plane \planp.
In fact only $\Omega_M$ is affected by a bias. \\

A weak $\Omega_M$ prior of
$\sigma (\Omega_M )=0.1$ has been used to draw Fig.~\ref{fig:scp}b, 
giving the Validity Zone for $w_0$ (identical to the full VZ).
The VZ is slightly reduced compared to the no prior case but it
 remains quite important within the plane \planp.
More precisely, it appears that the upper part of the biased zones are
comparable for both no and weak prior cases. Only the lower part of the
biased zone of Fig.~\ref{fig:scp}b is specific to the weak prior case. Note that changing
the error value of the prior has little consequences on the shape of the VZ.\\

This fitting procedure is
widely used to extract informations from present SN data \cite{SCP,Riess,Tonry,newSCP,Melchiorri}.
The results published in these references are valid for models with constant EoS. 
From our findings, one can extend the validity of these results and conclude that models
with a redshift dependent EoS which verify $\mid w_1^F \mid<0.7$ 
may be fitted by this 3-fit procedure with a bias which stays
 below the statistical error. \\
So, we conclude that we can
trust (at $1\sigma$) the results based on present SN data obtained with a 3-fit where we neglect
the redshift dependence of the EoS, using a $\Omega_M$ prior or not. Only (fiducial) models
with a strong recent variations of the EoS
may lead to biased results.\\

The comparison of the potentialities of the 3-fit and the 4-fit to
extract $w_0$ does not seem very relevant. However, we have checked this point, and it
appears that the 3-fit provides some constraints on $w_0$ which are roughly 30\%
better than with the 4-fit for most of \planp.\\

Finally, we have remarked that the central values of $w_0$ and
$\Omega_M$ are shifted as follows: if $w_1^F$ is positive we have
$w_0>w_0^F$ and $\Omega_M>\Omega_M^F$, and the converse if $w_1^F<0$.
We do not recover this simple behaviour with the SNAP and SNLS
samples.\\

\section{Summary and conclusions}

In this paper, we have studied the best strategy to extract the cosmological 
parameters and in particular the dark energy component from
the future supernovae data within the $\Omega_T=1$ paradigm. 
We have compared results from the present statistics with
the ones we can expect from future large ground-based surveys and from a future space
experiment like SNAP.\\
In order to be as model independent as possible we have looked at a large range of fiducial models
by varying the parameters of the equation of state of the dark energy 
$w_0^F$ and $w_1^F$ where $w^F(z)=w_0^F+w_1^Fz$ \cite{WA}.

With present statistics or the one expected from a survey like SNLS, 
precision on $w_1$ is very weak
and the physical question concerns the extraction of $w_0$ and $\Omega_M$.
For a SNAP like experiment a good precision on $w_1$ is only possible with 
a strong prior on $\Omega_M$.

To avoid the use of strong priors on $\Omega_M$, we have focused on the extraction of $w_0$ 
only and neglected the $w_1$ contribution by fixing it in the fitting procedure. We have to face
in this case the problem of introducing a bias if the fiducial value of $w_1^F$ is far away 
from the fixed one. We compare the bias to the statistical error which is expected
for each survey. 

To study the validity of such a strategy, we have presented an extensive study of fiducial models
in a flat universe in which we vary $w_0^F$ and $w_1^F$ in a complete range $-2<w_0^F<0$ 
and $-2<w_1^F<2$ corresponding to 
a large variety of DE theoretical models. 
On the other hand $M_s^F$ and $\Omega_M^F$ are fixed.
In the fit we have used in general a weak prior on $\Omega_M 
(\sigma(\Omega_M)< 0.1)$  and we have
looked at the bias introduced by fixing $w_1=0$. We call
this procedure the 3-fit ($M_s, \Omega_M, w_0$) procedure.

For SNAP (and SNLS) like statistics, we obtain the following conclusions :
\begin{itemize}
\item $w_0$  is not biased for an important part of the scanned plane
($w_0^F, w_1^F$) which corresponds to  models describing a Universe which is
today in acceleration.
\item for $M_S$, we get the same conclusion but we emphasize that this parameter is so 
much constrained
 in the fitting procedure that a small bias has no visible effect.
\item $\Omega_M$ is strongly biased if the redshift dependence is large ($|w_1^F|>0.2$(0.6)).
\item If a strong prior on $\Omega_M$ is used, $w_0$ is biased
for most of the plane. 
Namely, the 3-fit procedure is relevant to
extract $w_0$ if only a weak $\Omega_M$ prior is used.
\end{itemize}

We have compared the accuracy of the results on the $w_0$ parameter
to the ones obtained if $w_1$ is left free as in a 4-fit 
($M_s, \Omega_M, w_0,w_1$) procedure.

We found that, in spite of the bias, the 3-fit procedure, where $w_1$ is fixed, can be used for 
SNAP and SNLS to test accelerating models and give better or equivalent results
than the simplest 4-fit where $w_1$ is completely free. This is no longer true if $\mid w_1^F\mid$
is large. For decelerating models the 4-fit is mandatory.
Note that, concerning the choice of the number of parameters to be fitted, the bayesian
method presented in \cite{Bridle}
seems promising.\\

These conclusions have been set using 
only the value $\Omega_M^F=0.3$ for the fiducial models. If the true value
is different 
the areas of the different Validity Zones increase if the contribution of
DE is smaller, namely if $\Omega_M^F$ is larger. Conversely, if $\Omega_M^F<0.3$,
the validity of the 3-fit is reduced.\\

Using the presently available statistics of the SCP collaboration \cite{newSCP},
it appears that the procedure where $w_1$ is fixed can be trusted
for almost all the fiducial models considered here. Only fiducial models
with strong recent variations of the EoS ($\mid w_1^F \mid>0.7$)
may lead to biased results.\\

Let us point out that other bias would be introduced, by 
a departure from the $\Omega_T=1$ paradigm in the fiducial model, or if the central value
of the $\Omega_M$ prior itself (used in the fitting procedure) is not the same as the fiducial value.
Indeed a small bias in the prior could induce some strong bias for the other cosmological parameters.
These bias will be studied in a future paper \cite{enpreparation}. \\

It has been advocated recently \cite{Wang} that it should be easier to
constrain the DE density $\rho_X(z)$ and its time derivative
$\rho^{\,\prime}_X(z)$, instead of constraining the equation of state $w(z)$.

These authors have simulated a large variety of fiducial models
in the range $-1.2 \leq w^F_0 \leq -0.5$ and $-1.5 \leq w^F_1 \leq 0.5$,
which is almost completely contained in
our Validity Zone for SNAP (see Fig.~\ref{fig:statplussyst}).
They interpret
physically parts of this plane in term of
increasing, decreasing or non-monotonic DE densities $\rho_X(z)$.

Their approach is interesting to answer
if the DE energy density is a constant or not.
Nevertheless, focusing on the equation of state
itself remains mandatory if, for instance, 
one wants to validate precisely the acceleration
since, in this case, a precise knowledge of $w_0$
(and also on $\Omega_M$ and $\Omega_X$) is necessary.
Finally, we would like to point out that
trying to answer any particular question on the nature of DE requires 
an investigation on the choice of the most relevant
fitting procedures.\\

\m 

\no {\bf Acknowledgments}\\

We wish to thank A. Mazure and the members of the 
Laboratoire d'Astrophysique de Marseille, for their support,
and also E. Linder,
R. Miquel and F. Steiner, for helpful discussions.
``Centre de Physique Th\'eorique'' is UMR 6207 - ``Unit\'e Mixte
de Recherche'' of CNRS and of the Universities ``de Provence'',
``de la M\'editerran\'ee'' and ``du Sud Toulon-Var''- Laboratory
affiliated to FRUMAM (FR 2291).
``Centre de Physique des Particules de Marseille'' is ``Unit\'e Mixte
de Recherche'' of CNRS/IN2P3 and of the University
``de la M\'editerran\'ee''.\\




\begin{thebibliography}{99}


\bibitem{SCP} S. Perlmutter et al. Astrophys. J. {\bf 517}, 565 (1999)

\bibitem{Riess} A.G. Riess et al., Astron. J. {\bf 116}, 1009 (1998)

\bibitem{Tonry} J. Tonry et al., Astrophys. J. {\bf 594}, 1 (2003).

\bibitem{newSCP} R.A. Knop et al. astro-ph/0309368

\bibitem{cfht} see e.g. 
http://cfht.hawai.edu/Science/CFHTLS-OLD/history$\_$2001.html

\bibitem{Efst2DF}
G. Efstathiou et al., Mon. Not. R. Astron Soc. {\bf 330}, L39 (2002).

\bibitem{Spergel} D.N. Spergel et al. (WMAP Collaboration) 
Astrophys. J. Suppl. {\bf 148}, 175 (2003).

\bibitem{SDSSWMAP} M. Tegmark et al., astro-ph/0310723

\bibitem{Scranton} R. Scranton et al. astro-ph/0307335

\bibitem{Bahcall} see e.g. N. Bahcall, J.P. Ostriker, S. Perlmutter and P.J. Steinhardt,
Science {\bf 284}, 1481 (1999) and refs. therein.

\bibitem{Peebles} P.J.P. Peebles and B. Ratra, Rev. Mod. Phys., {\bf 75}, 559 (2003)
and references therein.

\bibitem{SNLS} see e.g. http://snls.in2p3.fr

\bibitem{SNAPweb} see e.g. http://snap.lbl.gov

\bibitem{Kim} A.G. Kim et al., Mon. Not. R. Astron Soc. {\bf 347}, 909 (2004).

\bibitem{Maor1} I. Maor, R. Brustein and P.J. Steinhardt, Phys. Rev. Lett.
{\bf 86}, 6 (2001).

\bibitem{Maor2} I. Maor et al., Phys. Rev. {\bf D65}, 123003 (2002).

\bibitem{WA} J. Weller and A. Albrecht, Phys. Rev. Lett. {\bf 86}, 1939 (2001); 
Phys. Rev. {\bf D65}, 103512 (2002), and references therein.

\bibitem{HT} D. Huterer and M.S. Turner, Phys. Rev. {\bf D64}, 123527 (2001).

\bibitem{Goliath} M. Goliath et al., Astron.\& Astrophys. {\bf 380}, 6 (2001).

\bibitem{Gerke} B. F. Gerke and G. Efstathiou, Mon. Not. R. Astron Soc. {\bf 335}, 33 (2002).

\bibitem{LH} E. V. Linder and D. Huterer, Phys. Rev.{\bf D67}, 081303 (2003).

\bibitem{DiPietro} E. Di Pietro and J.-F. Claeskens, 
Mon. Not. R. Astron Soc. {\bf 341}, 1299 (2003).

\bibitem{SNfactory} W.M. Wood-Vasey et al., astro-ph/0401513  

\bibitem{geo} V. Sahni et al., JETP Lett. {\bf 77}, 201 (2003); U. Alam et al.,
Mon. Not. R. Astron Soc. {\bf 344}, 1057 (2003); M. Tegmark, Phys. Rev. {\bf D66}, 103507 (2002);
D. Huterer and G. Starkman,   Phys. Rev. Lett. {\bf 90}, 031301 (2003);
P.S. Corasaniti and E.J. Copeland, Phys. Rev. {\bf D67}, 063521 (2003).

\bibitem{LinderPRL} E. V. Linder, Phys. Rev. Lett. {\bf 90}, 091301 (2003).

\bibitem{Frieman} J. A. Frieman et al., Phys. Rev. Lett. {\bf 75}, 2077 (1995).

\bibitem{enpreparation} J.-M. Virey et al., in preparation.

\bibitem{Frampton} P.H. Frampton, astro-ph/0302007 and references therein.

\bibitem{Nesseris} S. Nesseris and L. Perivolaropoulos astro-ph/0401556
and references therein.

\bibitem{kosmoshow} Our simulation tool, the ``Kosmoshow'', has been
developped by A. Tilquin and is available upon request or directly
at\\
http://marwww.in2p3.fr/renoir/Kosmoshow.html.

\bibitem{Blanchard} A. Blanchard et al.,  Astron.\& Astrophys. {\bf 412}, 35 (2003);
T. Shanks astro-ph/0401409.

\bibitem{Melchiorri} A. Melchiorri et al., Phys. Rev. {\bf D68}, 043509 (2003); 
S. Hannestad and E. Mortsell, Phys. Rev. {\bf D66}, 063508 (2002);
R. Bean and A. Melchiorri, Phys. Rev. {\bf D65}, 041302 (2002).

\bibitem{Bridle} T.D. Saini, J. Weller and S.L. Bridle, astro-ph/0305526.

\bibitem{Wang} Y. Wang et al., astro-ph/0402080 and references therein.

%



 
\end{thebibliography}
\end{document}